\documentclass[aps,prb,twocolumn,superscriptaddress,floatfix]{revtex4}
\usepackage{psfig}
\usepackage{graphicx}
\usepackage{epsfig}
\bibstyle{apsrev.bib}

\begin{document}
\input epsf.sty

\title{The partially asymmetric zero range process with quenched disorder}

\author{R\'obert Juh\'asz}
 \email{juhasz@lusi.uni-sb.de} 
\affiliation{Fachrichtung Theoretische Physik, Universit\"at des
  Saarlandes, D-66041 Saarbr\"ucken, Germany}
\author{Ludger Santen}
 \email{santen@lusi.uni-sb.de}
\affiliation{Fachrichtung Theoretische Physik, Universit\"at des
  Saarlandes, D-66041 Saarbr\"ucken, Germany}
\author{Ferenc Igl\'oi}  \email{igloi@szfki.hu}
\affiliation{ Research Institute for Solid
State Physics and Optics, H-1525 Budapest, P.O.Box 49, Hungary}
\affiliation{ Institute of Theoretical Physics, Szeged University,
H-6720 Szeged, Hungary}

\date{\today}

\begin{abstract}
  We consider the one-dimensional partially asymmetric zero range
  process where the hopping rates as well as the easy direction of
  hopping are random variables.  For this type of disorder there is a
  condensation phenomena in the thermodynamic limit: the particles
  typically occupy one single site and the fraction of particles
  outside the condensate is vanishing. We use extreme value statistics
  and an asymptotically exact strong disorder renormalization group
  method to explore the properties of the steady state.  In a finite
  system of $L$ sites the current vanishes as $J \sim L^{-z}$, where
  the dynamical exponent, $z$, is exactly calculated. For $0<z<1$ the
  transport is realized by $N_a \sim L^{1-z}$ active particles, which
  move with a constant velocity, whereas for $z>1$ the transport is
  due to the anomalous diffusion of a single Brownian particle.
  Inactive particles are localized at a second special site and their
  number in rare realizations is macroscopic. The average density
  profile of inactive particles has a width of, $\xi \sim
  \delta^{-2}$, in terms of the asymmetry parameter, $\delta$.  In
  addition to this, we have investigated the approach to the steady
  state of the system through a coarsening process and found that the
  size of the condensate grows as $n_L \sim t^{1/(1+z)}$ for large
  times. For the unbiased model $z$ is formally infinite and the
  coarsening is logarithmically slow.
\end{abstract}

\maketitle

\newcommand{\bc}{\begin{center}}
\newcommand{\ec}{\end{center}}
\newcommand{\be}{\begin{equation}}
\newcommand{\ee}{\end{equation}}
\newcommand{\beqn}{\begin{eqnarray}}
\newcommand{\eeqn}{\end{eqnarray}}

\vskip 2cm
\section{Introduction}

%%%%%%%%%%%%%%%%%%%%%%%%%%%%%%%%%%%%%%%%%%%%%%%%%%%%%%%%%%%%%%%%%%%%%%%%%%%
%%%%%%%%%%%%%%%%%%%%%%%%%%%%%%%%%%%%%%%%%%%%%%%%%%%%%%%%%%%%%%%%%%%%%%%%%%%

The properties of interacting many particle systems can be strongly
affected by the  presence of quenched disorder. This in particular
true for systems of self-driven particles\cite{krug,im,barma}, where even pointlike
defects\cite{janowsky} are able to change the macroscopic properties of the
system. Often these defects cause phase separated states, a
phenomenon, which is known from jam formation at bottlenecks\cite{krug,barma}. 

Particular interesting features arise, if not only the amplitude
of the hopping rates are quenched random variables, but the
directional bias as well. Then the dynamics of the particles is
governed by a complex landscape of energy barriers. As the escape
time growth exponentially with the heights of the barriers, the
largest barriers in the system determine the velocity of the
particles. This property of the many particle system is in agreement
with more classical problem of single particle diffusion in a
disordered environment\cite{sinai}, which is rather well understood and
serves for numerous applications, e.g. polymer translocation through a
narrow pore\cite{kln04} or the motion of molecular motors
 on heterogeneous tracks\cite{kln05}.

Since there is no general framework of studying nonequilibrium
disordered systems it is of interest to investigate specific simple
models.  Here we consider the zero range process\cite{evansreview} (ZRP) with quenched
disorder. The ZRP is particularly well suited for theoretical analysis
because the stationary weights of a given configuration factorize and
can be exactly calculated\cite{spitzer}. But the ZRP is not only conceptually
interesting, there are a number of important applications as well: it
has been used in order to describe e.g. the formation of traffic jams\cite{chowd},
the coalescence in granular systems, and the gelation in networks.

The effect of quenched disorder on the properties of the ZRP have been
investigated in one dimension in the totally asymmetric version\cite{jainbarma}, i.e. when the
particles hop in one direction with position dependent rates. In this
case a dynamical phase transition takes place from a low density
phase, where one observes the condensation of holes, to a homogeneous
high density state\cite{kf,evans}.  At low densities the average speed of the
particles is related to the lower cut-off of the effective hopping
rates. The properties of the phase transition are determined by the
asymptotics of the effective hopping rate distribution at the lower
cut-off and not by its mean or variance.  If the ZRP is partially
asymmetric and the easy direction of hopping is random as well, one
observes a strong dependence of the average speed of the particles on
the system size rather than on the density as in case of the totally
asymmetric model\cite{jsi}. Already this result illustrates the 
qualitative difference
between the two different realization of the disorder.

The ZRP can be exactly mapped 
onto a one-dimensional asymmetric simple
exclusion process\cite{liggett,hinrichsen,schutzreview} (ASEP) if sites are considered as particles and
masses as hole clusters. Disorder in the ZRP is transformed into
particle dependent hopping rates in the ASEP, which is generally
referred to as particle-wise disorder. In the ASEP one can also realize disorder
which depends on sites. The ASEP both with particle-wise\cite{kf,evans} and
site-wise disorder\cite{barma,derrida,stinchcombe,jsi} have been extensively studied, in particular for systems
which are totally asymmetric. 

The partially asymmetric ASEP with particle-wise disorder has been recently
studied by a strong disorder
renormalization group (RG) method\cite{jsi} which provides asymptotically
exact results for large sizes. This RG
method has originally been introduced in order to
study random quantum spin chains\cite{MDH}, but afterwards it has been applied
for a large variety of quantum\cite{DF,fisherxx} as well as classical\cite{RGsinai} systems, both in
and out of equilibrium\cite{hiv}, for a review see\cite{im}.

In this paper we use this RG method to study the
partially asymmetric ZRP with quenched disorder. With this type of
disorder there is a condensation phenomenon: almost all particles
occupy one single site, whereas the fraction of particles outside the
condensate converges to zero in the thermodynamic limit. Here we address
questions regarding the size dependence of different quantities, such
as the stationary current, the density profile, the mean density in
the bulk and the number of particles outside the condensate. We also
investigate the approach to the steady state through a coarsening
process and the time dependence of the mass of the condensate.

The paper is organized as follows. In the next section we will
introduce the disordered ZRP and discuss its relation to the ASEP with
particle disorder. In Sec.\ref{sec_RG} we introduce the
RG method which is used to calculate the steady
state current in the system. The density profile as well as
finite-size behavior of the bulk density is calculated in
Sec.\ref{sec_profile}, whereas the coarsening behavior is analyzed
in Sec.\ref{sec_coarsening}.  We summarize and discuss our results in
the final section. Some details of the calculations are given in  the
Appendices.

%%%%%%%%%%%%%%%%%%%%%%%%%%%%%%%%%%%%%%%%%%%%%%%%%%%%%%%%
\section{Definitions and preliminaries} 
\label{model}
%%%%%%%%%%%%%%%%%%%%%%%%%%%%%%%%%%%%%%%%%%%%%%%%%%%%%%%%

%%%%%%%%%%%%%%%%%%%%%%%%%%%%%%%%%%%%%%%%%%%%%%%%%%%%%%%%%%%%%%%%%%%%%%%%%%%
\begin{figure}
\includegraphics[width=0.8\linewidth]{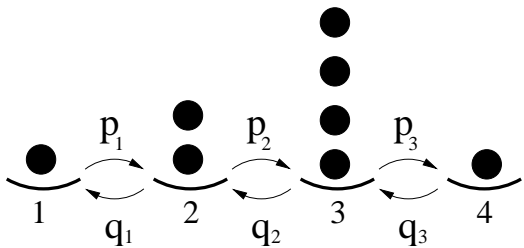}
\caption{\label{fig_zrp} The zero range process on a one-dimensional
  lattice. The top most particle at site $i$ hops with rate $p_i$ to
  site $i+1$ and with rate $q_{i-1}$ to site $i-1$.}
\end{figure}
%%%%%%%%%%%%%%%%%%%%%%%%%%%%%%%%%%%%%%%%%%%%%%%%%%%%%%%%%%%%%%%%%%%%%%%%%%%

In the ZRP particles hop from site to site on a lattice and the hop
rates depend on the departure site and on the number of particles
(mass) at that site. With these conditions the stationary weight of a
configuration is given in a factorized form, which offers an
opportunity to analyze the steady state properties exactly\cite{evansreview}. 
The type of ZRP we consider in the present work
is defined on a one-dimensional periodic lattice with
$l=1,2, \dots,L$ sites and $N$ particles. 
Particles are allowed to hop to nearest neighbor sites. 
The hopping rates are quenched random variables,
where we denote
the forward hopping rate from site $l$ to $l+1$  by $p_l$ and the
backward hopping rate from site $l+1$ to $l$ by $q_l$ (see Fig.~\ref{fig_zrp}).
Compared to the most general ZRP, we do not consider mass-dependent
hopping rates, i.e. $p_l$ and $q_l$ are valid for all occupation numbers 
$n_l\geq 1$.
A configuration of the system is characterized by the distribution
of masses (number of particles), $n_l$, where $\sum_{l=1}^L n_l=N$.

The one-dimensional ZRP defined above is equivalent to an ASEP where
$l=1,2, \dots, L$ particles are placed on a ring with $L+N$
sites. The $l$-th particle hops to empty neighboring sites with a
forward (backward) rate of $q_l$ ($p_l$) and a configuration is
defined by the number of empty sites, $n_l$, behind the particle (i.e.
in front of the particle there are $n_{l+1}$ holes).

The stationary weight of a configuration of the ZRP is
given in a product form:
\be
P(\{n_l\})  =  Z^{-1}_{L,N} \prod_{l=1}^L  f_l(n_l)
\label{product}  
\ee
where
\be 
Z_{L,N}  = \sum_{\{n_l\}} \prod_{l=1}^L f_l(n_l)\delta \left  ( \sum_{l=1}^{L}  n_l - N  \right )\;
\label{ZLN}
\ee
is the canonical partition sum. The factors are given by:
\be
f_l(n_l)=g_l^{n_l}\;
\ee
and the $g_l$ satisfy the equations:
\be
g_l(p_l+q_{l-1})=g_{l-1} p_{l-1} + g_{l+1} q_l\;.
\label{geq}
\ee
These equations are identical to the stationary weights of a random
walker with random hop rates: $p_l$ ($l \to l+1$) and $q_{l-1}$ ($l
\to l-1$). The solution of Eq.~(\ref{geq}) is given by:
\be
g_l=\frac{C}{p_l} \left[ 1 + \sum_{i=1}^{L-1} \prod_{j=1}^i \frac{q_{l+j-1}}{p_{l+j}} \right]\;,
\label{g}
\ee
where $C$ is a constant. From Eq.~(\ref{geq}) one obtains:
\be
g_{l-1} p_{l-1} - g_l q_{l-1}=g_l p_l - g_{l+1} q_l=const\;,
\label{const}
\ee
which is a constant of motion. Taking the constant in Eq.~(\ref{const})
to be one we get:
\be
C=\left[ 1 - \prod_{l=1}^L \frac{q_l}{p_l} \right]^{-1}\;.
\ee
One can obtain a number of useful results for the stationary 
state\cite{evansreview}.
E.g. the occupation probability, $p_l(n_l)$, that the $l$-th
site contains $n_l$ particles is given by:
\be
p_l(n_l)=f_l(n_l) \frac{Z_{L-1,N-n_l}} {Z_{L,N}} \;,
\label{p_n}
\ee
and the particle current reads as:
\be
J_{L,N}=\langle p - q \rangle
=\frac{Z_{L,N-1}}{Z_{L,N}}\;.
\label{J}
\ee

These expressions are simplified if the number of particles goes to
infinity, $N \to \infty$, whereas there is no restriction on the value
of $L$. In this limit we obtain for the canonical partition sum:
\be
Z_{L,N}=g_L^N \prod_{l=1}^{L-1} \frac{1}{1-g_l/g_L},\quad N \to \infty\;,
\label{Z_inf}
\ee
where we label the sites in a way that $g_L=\max(\{g_l\})$. Then the
stationary current reads as:
\be
J_{L}=1/g_L,\quad N \to \infty\;,
\label{JL}
\ee
and the occupation probability
follows a geometrical distribution:
\be  
p(n_l)=(1-\alpha_l)\alpha_l^{n_l},  \quad \alpha_l=g_l/g_L, \quad N \to \infty\;,
\label{geom}
\ee
so that
\be
\langle n_l \rangle =\alpha_l/(1-\alpha_l), \quad N \to \infty.
\label{nav}
\ee 

For large, but finite $N$ and $L$ the sum of the occupation numbers
should be equal to the total number of particles\cite{evansreview}:
\be \sum_{l=1}^{L}\langle n_l
\rangle=\sum_{l=1}^{L}{1\over 1/(g_lJ_{L,N})-1} =N\;.
\label{finite}
\ee

In this work the hop rates are independent and identically
distributed random variables taken from the distributions, $\rho
(p)dp$ and $\pi (q)dq$, respectively which will be specified later.
Generally we allow for the existence of links with $p_l>q_l$ as well
as links with $p_l<q_l$ with finite probability, i.e. the easy
direction of hopping is a random variable, too.

We introduce a control-parameter, $\delta$,
which characterizes the average asymmetry between forward and backward rates:
\be \delta=\frac{[\ln  p]_{\rm av}  - [\ln q]_{\rm  av}}{{\rm var}[\ln
p]+{\rm var}[\ln q]}\;,
\label{delta}
\ee
such that for $\delta>0$ ($\delta<0$) the particles move on average to
the right (left).  Here, and in the following $[\dots]_{\rm av}$
denotes average over quenched disorder, whereas ${\rm var}(x)$ stands
for the variance of $x$.

One can show, that for any non-zero density, $\rho=N/L>0$, a Bose
condensation occurs, in the sense
that a finite fraction of the particles,
$\langle n_L \rangle/N >0$, are condensed at site $L$, and $\langle
n_L \rangle/N$ tends to one in typical samples for $N \to \infty$.  
The condensation of particles can be understood by analyzing
the distribution of  $g_i$'s, as we will show in the next section  
(see for comparison Refs.[\onlinecite{evans,jsi}]). 
We shall also show that the stationary
current vanishes in the thermodynamic limit, i.e. $\lim_{L\to\infty}J_L=0$.
These results are in complete agreement with a recently introduced
criterion \cite{kafri}, which predicts strong phase separation for
vanishing stationary current.

The numerical calculations are carried out using two kinds of disorder 
distributions. First, a bimodal distribution, where 
$p_iq_i=r$ holds for all $i$ and \be \rho (p)=c\delta (p-1)+(1-c)\delta
(p-r),
\label{bimodal}
\ee
with $r>1$ and $0<c\le 1/2$.
Second, we use a uniform distribution defined by:
\beqn
\rho (p)&=&p_0^{-1}\Theta (p)\Theta (p_0-p), \nonumber \\
\pi (q) &=& \Theta (q)\Theta (1-q),
\label{uniform}
\eeqn
where $p_0>0$ and $\Theta (x)$ is the Heaviside function.

In the following we use the terminology, that the
model is asymmetric or biased, for $\delta>0$, and it is unbiased for
$\delta=0$. This latter model is realized  if the distribution of the
hopping rates is symmetric. For random quantum spin chains, which show analogous
low-energy properties, $\delta=0$, corresponds to the critical point and
(a part of) the $\delta>0$ region is the so called Griffiths phase\cite{griffiths}.
Since the same type of mechanism takes place in the two systems we use the
terminology of Griffiths phase for the random ZRP with $\delta>0$, too.

%%%%%%%%%%%%%%%%%%%%%%%%%%%%%%%%%%%%%%%%%%%%%%%%%%%%%%%%%%%%%%%%%%%%%%%%
\section{Analysis of the current}
\label{sec_current}

In section~\ref{model} we pointed out that the stationary 
weights of the disordered ZRP are
related to the statistical weights, $g_l$, of a random walker in a 
random environment. 
For the further analysis it is important to notice that the
random variables $g$ defined by Eq.(\ref{g}),
are so-called Kesten variables\cite{kesten}, because there exist a number of rigorous 
mathematical results concerning their distribution $P_L(g)$. In the 
Griffiths phase, i.e. for
$\delta>0$ and in the thermodynamic limit, $L \to \infty$, the limit
distribution presents an algebraic tail:
\be
P_{L}(g) \sim g^{-1-1/z}, \quad L \to \infty\;,
\label{kesten}
\ee
where $z$ is the positive root of the equation
\be 
\left[\left({q\over p}\right)^{1/z} \right]_{av}=1.
\label{z}
\ee
In particular we obtain for small, $\delta$, that:
\be
z \approx \frac{1}{2 \delta}, \quad \delta \ll 1\;,
\label{z_small}
\ee
which is divergent and independent of the actual form of the
distributions, $\rho(p)$ and $\pi(q)$.

If we now apply the results for Kesten variables given above
to the disordered ZRP two remarks are in order.
First, the algebraic tail in Eq.~(\ref{kesten}) should be present
in the large (but finite) $L$ limit. Second, the $g_l$-s in
Eq.~(\ref{g}) are not
strictly independent. However, as will be shown
later $g_l$ and $g_k$ can be considered to be uncorrelated
if  $|l-k|>\xi$, where $\xi$ is
the finite correlation length of the problem. Therefore
$g_L$ has to be treated 
as the largest event of a distribution in
Eq.~(\ref{kesten}) among $\sim L/\xi$ terms. According to extreme value
statistics\cite{galambos} the typical value of $g_L$ follows from the relation:
$g_L^{-1/z}L=O(1)$, such that:
\be
g_L \sim L^{z}\;.
\label{g_L}
\ee
Another result, which we obtain from extreme value statistics, 
is the typical value of the $i$-th largest occupation probability, $g(i)/g_L$:
\be
\frac{g(i)}{g_L} \approx i^{-z}\;.
\label{g_i}
\ee
From these results and from Eq.~(\ref{J}) follows that the typical
value of the current is
\be 
|J_L|\sim L^{-z},\quad \delta>0\;,
\label{jgriffiths}
\ee
thus vanishes algebraically in the thermodynamic limit. Our second
result concerns the typical value of particles outside the condensate,
$N_{out}$, which follows from Eqs.(\ref{nav}) and (\ref{g_i}) as:
\be
N_{out} = \sum_{l=1}^{L-1} \langle n_l \rangle \sim \sum_{i=1}^{L-1} i^{-z}
\sim \left\{ \begin{array}{l} L^{1-z},\quad z<1 \\
                           O(1), \quad z>1  \end{array} \right. \;.
\label{N_out}
\ee
Therefore, for any $z>0$  the fraction of
particles in the condensate goes to one in the limit of large 
system sizes $L\to\infty$ and constant density, $\rho=L/N$. 
The typical number of particles outside the condensate, however,
 behaves differently for $z<1$, when
it is divergent, and for $z>1$, when it tends to a finite value. We
shall discuss this issue in more detail in Sec.\ref{sec_profile}.

\begin{figure}
\includegraphics[width=0.8\linewidth]{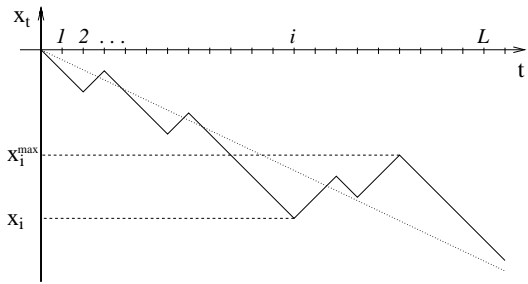}
\caption{\label{rw1} Mapping the configurations to random walk paths.}  
\end{figure}

Next, we turn to analyze the behavior of the $g_l$ weights in the
weakly asymmetric limit, $\delta \ll 1$ and estimate the correlation
length, $\xi$. 
For our analysis it is convenient to use a correspondence between the
sequences of the hop rates and random walk paths. (A similar reasoning has been introduced for
the random transverse-field Ising chain in Ref.\cite{ir}.) 
We consider the situation where the walker starts
at $i=0$, $x_0=0$ and takes in its
$l$-th step an oriented distance of $\delta x_l=\ln q_{l-1} - \ln
p_l$, such that its position, $x_l$, is given by: $x_l=\sum_{i_1}^l
\delta x_i$, see Fig. \ref{rw1}. For $\delta > 0$ the
motion of the walker is biased (due to the average slope of the 
energy landscape), but
decorated by fluctuations, i.e. one observes local deviations 
from the average behavior. The typical time during which the walker makes a large
excursion, $\xi$, against the bias follows from the Gaussian nature of the fluctuations
and given by\cite{ir}:
\be
\xi \sim \delta^{-2}\;.
\label{xi}
\ee
More precisely the probability of an excursion time, $i$, reads as
$P(i) \sim \exp(-i/\xi)$.  Similarly, from the Gaussian form of the
fluctuations follows the typical transversal size of excursions of
the walker, which scales as, $\xi_{\perp} \sim \xi^{1/2} \sim \delta^{-1}$.
Consequently the probability of a transversal size of excursions,
$\Delta$, is given by $P(\Delta) \sim \exp(-\Delta/\xi_{\perp})$.

$\xi$ as defined in Eq.~(\ref{xi}) is the correlation length of the
random walk and at the same time it is the correlation length of the
ZRP as discussed in the first part of this section. Indeed, the
stationary weight, $g_i$, is connected to the part of the landscape
which starts at position, $i$, and its value is dominated by the height of the
largest excursion, $x_i^{max}-x_i$, see Fig.\ref{rw1}. Since the
landscape becomes uncorrelated for distances (times), which are larger
than $\xi$, the corresponding $g_l$ weights are
uncorrelated, too. The value of $g_L$, i.e. the largest weight, is
related to the largest possible transverse fluctuation, $\Delta_L$. In
a chain of length, $L$, this extremal position can be chosen out of
$\sim L$ positions, therefore the typical value of $\Delta_L$ follows
from extreme value statistics as: $\exp(-\Delta_L/\xi_{\perp}) L
=O(1)$, thus $\Delta_L \sim \ln L \delta^{-1}$. Keeping in mind the
definition of $\delta x_i$ and putting this result into Eq.~(\ref{g})
we obtain: $\ln g_L \sim \delta^{-1} \ln L$, which is compatible with
the exact result in Eq.~(\ref{g_L}) and the small $\delta$
expansion of $z$ in Eq.~(\ref{z_small}).

In the random unbiased ZRP $\delta=0$ and the correlation length is
divergent. In this case the transverse fluctuations of the walker in
the thermodynamical limit are unbounded. In a finite system
they typically behave as: $\Delta_L \sim L^{1/2}$. 
From this follows that
the fluctuations in the particle current in a typical sample are of
the form:
\be 
|\ln |J_L|| \sim L^{1/2},\quad \delta=0\;.
\label{jcrit}
\ee

We can thus conclude that in the random walk picture the largest local
fluctuation of the energy landscape is responsible for the small value of the
particle current. The particles are accumulated in front of this large
barrier and built the condensate at a given site of the system.  
At other subleading barriers there are only a few, $O(1)$, particles
the sum of which gives typically $N_{out}$, as estimated in
Eq.~(\ref{N_out}). Later in Sec.\ref{sec_cloud} and \ref{sec_density}
we shall use this random walk mapping to obtain the size of the cloud
of particles around the condensate and the density profile.

%%%%%%%%%%%%%%%%%%%%%%%%%%%%%%%%%%%%%%%%%%%%%%%%%%%%%%%%%%%%%%%%%%%%%%%%%%%
%%%%%%%%%%%%%%%%%%%%%%%%%%%%%%%%%%%%%%%%%%%%%%%%%%%%%%%%%%%%%%%%%%%%%%%%%%%

%%%%%%%%%%%%%%%%%%%%%%%%%%%%%%%%%%%%%%%%%%%%%%%%%%%%%%%%%%%%%%%%%%%%%%%%%%%%%%%%%%
\section{Renormalization and the particle flow}
\label{sec_RG}
The results about the particle current presented in the previous
section can be obtained, together with another results, by the
application of an RG method. Here we first
illustrate how single sites of the lattice can be decimated,
afterwards the method is adopted to the random system and finally the
properties of the fixed points, both for the unbiased and asymmetric (biased)
models are discussed.
%%%%%%%%%%%%%%%%%%%%%%%%%%%%%%%%%%%%%%%%%%%%%%%%%%%%%%%%%%%%%%%%%%%%%%%%%%%
\begin{figure}
\includegraphics[width=\linewidth]{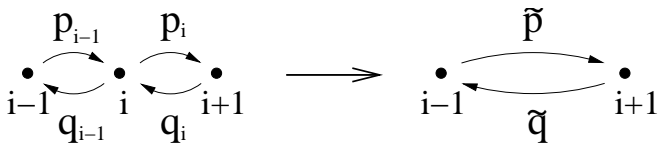}
\caption{\label{fig_elim} Renormalization scheme for the zero range
  process: Site $i$ is eliminated and
  hopping from site $i-1$ to $i+1$ and vice versa occurs with 
  rates $\tilde p$ and $\tilde q$, respectively.}
\end{figure}
%%%%%%%%%%%%%%%%%%%%%%%%%%%%%%%%%%%%%%%%%%%%%%%%%%%%%%%%%%%%%%%%%%%%%%%%%%%

\subsection{Exact decimation of a single site}

The factorized form of the stationary distribution in
Eq.~(\ref{product}) allows to explicitly integrate out a
typical site, say $i \ne L$. To do this we start with Eq.~(\ref{const})
and eliminate $g_i$: \be {p_{i-1}p_{i}\over
  p_i+q_{i-1}}g_{i-1}-{q_{i-1}q_{i}\over p_i+q_{i-1}}g_{i+1} = 1, \ee
so that the effective (two site) hop rates are identified as
\be 
\tilde p = {p_{i-1}p_{i}\over p_i+q_{i-1}}, \qquad  
\tilde q = {q_{i-1}q_{i}\over p_i+q_{i-1}}.
\label{effective}
\ee
At the same time the particle current is transformed as:
\be
\tilde{J}=\sum_{n_i=0}^{\infty} \frac{Z_{L-1,N-n_l-1}}{Z_{L-1,N-n_l}} p(n_l)\;,
\label{Jtilde}
\ee
what can be obtained along the lines of Eq.~(\ref{J}). From a practical
point of view it is of importance that the particle current remains
invariant, which happens both, i) for $N \to \infty$ (even for finite $L$)
and ii) if $\langle n_i \rangle \to 0$. For a random system case ii)
is realized for a typical site, if $L \to \infty$, even if
$N$ is finite. Indeed, 
in this case $q_i=O(1)$ whereas
$g_L=J_{L,N}^{-1} \to \infty$.  Since a site outside the condensate
generally contains finite number of particles these are all
``typical'' so that one expects to be able to repeat the
transformation up to site, $L$.

\subsection{Renormalization of the random ZRP}

In the following we shall apply the RG transformation in case ii),
i.e. for a random system in the infinite (or very large) lattice
limit. The corresponding single particle problem of a random
walker in a random potential is thoroughly studied in the literature
and many exact results have been obtained by using the strong disorder RG
method\cite{RGsinai,im}. The principles of the RG procedure
apply as well to the
the ZRP with many particles.
There are, however, questions which
are specific in the large $N$ limit, such as the properties of the
condensate, the profile of the particles and the coarsening process
towards the stationary state, which will be studied in the following
sections.

In the following we adopt the RG rules in Eq.~(\ref{effective}) for the
random system. In the traditional application of the strong disorder
RG method for a Brownian particle one renormalizes the energy
landscape\cite{RGsinai} in Fig.~\ref{rw1}. Here we apply a somewhat different
approach and also refer to a mapping to random quantum spin chains.
The first step is to select the site (bond) to be decimated out.
We choose the fastest rate, $\Omega= \max(\{p_i\},\{q_i\})$, which
means that processes which are faster than the time-scale,
$\tau=1/\Omega$, are integrated out.  If the distribution of disorder
is sufficiently broad, then the rate $\Omega$ is much larger than
the neighboring rates and the expressions for the renormalized rates
in Eq.~(\ref{effective}) can be simplified. For example if $\Omega=p_i$
we have:
\be 
\tilde p \approx p_{i-1}, \qquad
\tilde q \approx {q_{i-1}q_{i}\over \Omega },
\label{back}
\ee
and similarly for $\Omega=q_{i-1}$, by replacing $q_i$ and $p_i$. It
is evident from Eq.~(\ref{back}) that the generated new rate is smaller
than the eliminated rate $\Omega$. Repeating the decimation
procedure the energy scale is gradually lowered and we monitor the
distribution of the hop rates, $P(p,\Omega)$ and $R(q,\Omega)$.
respectively. In particular we are interested in the scaling
properties of the transformation at the fixed point, which is located
at $\Omega^*=0$. This latter statement is in accordance with the fact,
that the stationary current vanishes in the system.

Using the approximate decimation rules in Eq.~(\ref{back}) the RG
equations can be formulated in the continuum limit as a set of
integral-differential equations, that can be exactly solved at the
fixed point, both at $\delta =0$ \cite{DF} and for $\delta\neq0$
\cite{i02}. These calculations are identical to that of the random
transverse-field Ising chain (RTFIC) and we borrow the results
obtained for the latter model.  Details about the mapping between the
two problems as well as the fixed-point solutions are presented in
Appendix A.

\subsection{The unbiased ZRP}

The unbiased ZRP with $\delta =0$ corresponds to the critical point
of the RTFIC. At this point distribution of forward and backward
rates in Eq.~(\ref{sol}) are identical having the same exponents:
$r_0=p_0\sim 1/\ln(\Omega/\Omega_0)$. Here, $\Omega_0$ is a reference
energy scale. The appropriate scaling variable is given by $\eta=-(\ln
\Omega - \ln p)/\ln \Omega=-(\ln \Omega - \ln q)/\ln \Omega$, having a
distribution $\rho(\eta)=\exp(-\eta) {\rm d} \eta$,  $\eta > 0$. The
length scale, $L_{\Omega}$, which is the size of the effective
cluster, and the energy scale are related as:
\be
L_{\Omega} \sim \left[ \ln \frac{\Omega_0}{\Omega} \right]^2,\quad \delta=0\;,
\label{scale_crit}
\ee
which shows unusual, activated scaling. In the ZRP in the course 
of renormalization
huge particle clusters are created and the
distance, $X$, they travel is the accumulated distance covered by the
original particles that form the cluster, $X=\sum_{k=1}
x_k$. Finally after eliminating all but the last site, we
obtain the accumulated distance traveled by all the particles during
time, $t$.  In the unbiased case $[\langle X \rangle ]_{\rm av}=0$
holds and the average mean-square of the accumulated displacement is given by:
\be
[\langle X^2 \rangle ]_{\rm av} \sim \ln^4 t\;,
\label{sinai}
\ee
in agreement with the diffusion of a Sinai walker\cite{sinai1}. 
Furthermore an
appropriate scaling combination between current (measured in a
specific sample) and size is obtained as: $|\ln
|J_L|| L^{-1/2}$, which is compatible with the relation in
Eq.~(\ref{jcrit}).

\subsection{The asymmetric ZRP}
\label{sec_asym}

The ZRP with a global bias, $\delta >0$ (or equivalently $\delta <0$),
corresponds to the so called disordered Griffiths phase
\cite{griffiths} of the RTFIC. In this case a time scale $\tau 
\sim\Omega_{\xi}$ exists, which separates two characteristic 
areas of the renormalization process.
In the initial part of
the renormalization, for $\Omega>\Omega_{\xi}$, both backward and
forward rates are decimated out, until effective clusters of typical
size, $\xi$, are created, where the correlation length $\xi$ is defined in
Eq.~(\ref{xi}), through Eq.~(\ref{scale_crit}). 
 In the random walk picture in Fig. \ref{rw1}
this corresponds to eliminate the fluctuations of the landscape and
obtain a monotonically decreasing curve.  Now continuing the decimations
for $\Omega<\Omega_{\xi}$, almost exclusively forward rates are
eliminated and the ratio of the typical backward and forward rates
tends to zero. Consequently the system is renormalized to a totally
asymmetric ZRP in which the distribution of the forward rates can be
calculated (see Eq.~(\ref{sol})), and follows a pure power-law (see
Eq.~(\ref{sol})):
\be
P_0(p,\Omega) \approx \frac{1}{z \Omega}\left(\frac{\Omega}{p}\right)^{1-1/z},
\quad \Omega < \Omega_{\xi}\;,
\label{Pp}
\ee
where $z$ is the dynamical exponent as defined in Eq.~(\ref{z}). 
For comparison with Eq.~(\ref{kesten}) one should note that for the
totally asymmetric model the weights in Eq.~(\ref{g}) are $\sim 1/p_l$,
consequently Eqs.~(\ref{Pp}) and (\ref{kesten}) are of  identical form.
Furthermore the effective particles become indeed uncorrelated in a
length-scale of $\xi$.

If effective clusters of size $L$ are created the corresponding energy
scale is lowered as:
\be
L \sim \left(\frac{\Omega}{\Omega_0} \right)^{-1/z}, \quad \delta > 0\;.
\label{scale_gr}
\ee
Consequently the smallest effective forward hop rate is given by:
$\tilde{p}_L \sim L^{-z}$.  This implies the same relation for the
stationary current as given in Eq.~(\ref{jgriffiths}).

The results presented in the previous subsections are expected to be
asymptotically exact. Indeed during renormalization the distributions
of the hop rates broaden without limits both at the critical point and
in the Griffiths phase consequently the decimation rule in
(\ref{back}) becomes exact in the fixed point.

\begin{figure}
\includegraphics[width=1.0\linewidth]{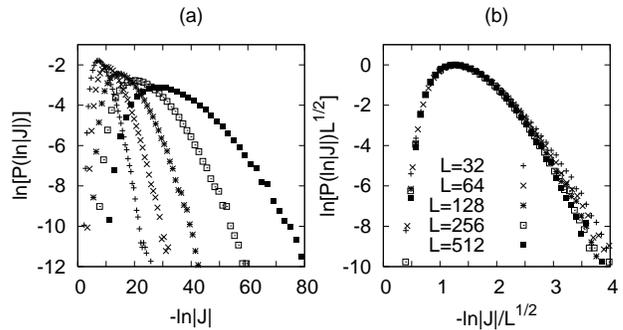}
\caption{\label{du1} Distribution of the steady state current in the random
  unbiased ZRP calculated from Eqs.(\ref{g}) and (\ref{JL}). a) Note
  that the distribution broadens with the size, which is a clear
  indication of an infinite disorder fixed point. b) An appropriate
  scaling collapse is obtained using the scaling combination in
  Eq.~(\ref{jcrit}).}
\end{figure}

\begin{figure}
\includegraphics[width=1.0\linewidth]{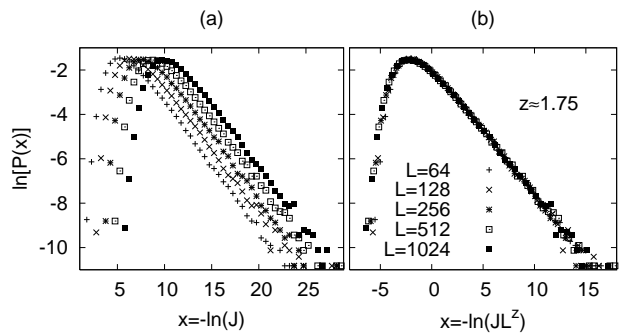}
\caption{\label{du2}  Distribution of the steady state current in the random
  asymmetric ZRP, i.e. in the Griffiths phase calculated from
  Eqs.(\ref{g}) and (\ref{JL}). a) In a log-log plot the distribution
  is shifted to the left with increasing sizes, the asymptotic slope
  of the curves is $1/z$ and can be used to measure the dynamical
  exponent.  b) An appropriate scaling collapse is obtained using the
  scaling combination in Eq.~(\ref{jgriffiths}) and the exact result
  for $z$ in Eq.~(\ref{z}). }
\end{figure}

The scaling forms of the current in Eqs.(\ref{jgriffiths}) and
(\ref{jcrit}) are tested by numerical calculations in Ref.\cite{jsi}
for the partially asymmetric ASEP with bimodal, i.e. discrete disorder
in Eq.~(\ref{bimodal}). Here we performed calculations on the ZRP using
the uniform, i.e. continuous disorder in Eq.~(\ref{uniform}). The
agreement is indeed satisfactory, both for the unbiased,
(Fig.~\ref{du1}) and for the biased (asymmetric) (Fig.~\ref{du2}) cases.

%%%%%%%%%%%%%%%%%%%%%%%%%%%%%%%%%%%%%%%%%%%%%%%%%%%%%%%%%%%%%%%%%%%%%%%%%%%
%%%%%%%%%%%%%%%%%%%%%%%%%%%%%%%%%%%%%%%%%%%%%%%%%%%%%%%%%%%%%%%%%%%%%%%%%%%

%%%%%%%%%%%%%%%%%%%%%%%%%%%%%%%%%%%%%%%%%%%%%%%%%%%%%%%%%%%%%%%%%%%%%%%%%%%
%%%%%%%%%%%%%%%%%%%%%%%%%%%%%%%%%%%%%%%%%%%%%%%%%%%%%%%%%%%%%%%%%%%%%%%%%%%

%%%%%%%%%%%%%%%%%%%%%%%%%%%%%%%%%%%%%%%%%%%%%%%%%%%%%%%%%%%%%%%%%%%%%%%%%%%
%%%%%%%%%%%%%%%%%%%%%%%%%%%%%%%%%%%%%%%%%%%%%%%%%%%%%%%%%%%%%%%%%%%%%%%%%%%
\section{Distribution of particles}
\label{sec_profile}

In order to characterize the microscopic states of the disordered 
ZRP it is instructive to distinguish between the
$\langle n_L \rangle$ particles in the condensate and
particles outside the condensate, which are either considered
to be {\it active}, i.e. they behave as a single random walker 
and carry the current of the system, or {\it inactive} particles 
that are localized outside the condensate.

\subsection{Active particles}

In a system with large finite number of sites, $L$, the number of
active particles is expected to scale as, $N_{a} \sim L^a$, with an
exponent $0 \le a <1$. The value of $a$ can be estimated by using the
relation: $J_{L}=v N_a/L$, where $v$ is the average velocity of a
Brownian particle in that landscape. In a finite system $v$ scales as:
$v \sim L^{-\omega}$, with $\omega=\max(z-1,0)$. Indeed a single
Brownian walker moves with a constant speed for $z<1$ and has
anomalous diffusion for $z>1$. Thus the particle current in a finite
system behaves as: $J_{L} \sim L^{a-1-\omega}$. This result should be
compared to our analysis in Eq.~(\ref{jgriffiths}), where we
distinguish between the  $z > 1$ and $z < 1$, respectively.

\subsubsection{Single particle transport  $z > 1$}

For $z>1$ we have $\omega=z-1$ and the current is so small that it is
produced by a finite number of active particles, i.e. $N_{a}=O(1)$. In
this case the accumulated distance traveled by all particles is simply
given by:
\be
X \sim t^{1/z}, \quad z>1\;.
\label{X_>}
\ee

\subsubsection{Many particle transport  $z < 1$}

For $z < 1$, however, $\omega=0$ and $a=1-z$, thus the current in the
ZRP is produced by $N_a \sim L^{1-z}$ active particles. 
Now the active
particles have a constant velocity, thus during time, $t$ they travel
a distance of $\sim t$.

Note that the number of active particles, $N_a$, is of the same order
of magnitude as the typical value of particles outside the condensate,
$N_{out}$, see Eq.~(\ref{N_out}). This is due to the fact that in a
typical sample there are only finite number of inactive particles,
this issue is discussed in details in the following subsection.

\subsection{Inactive particles}

Inactive particles in the random ZRP are also of two kinds. The first
kind of inactive particles is found in a ``cloud'', which is localized 
next to the condensate. The formation of
this cloud is due to an attractive property of the
energy landscape: particles that left the condensate
and did not travel farther than a
distance  $\xi$ will typically turn back.
The second sort of inactive particles are found to be localized at the
subleading extrema of the energy landscape, i.e. they are at a site
$\tilde{l}$, where $g_{\tilde{l}}$ in Eq.~(\ref{g}) takes the
next-to-leading value. As shown in Eq.~(\ref{g_i}) the typical value of
$\langle n_{\tilde{l}} \rangle$ is of the order of one. Its average
value, however, is divergent and will be determined below.

\subsubsection{Typical behavior of the cloud}
\label{sec_cloud}

Here we consider a typical sample in which the cloud is due to the
attraction of the condensate and well separated from
the subleading extrema of the landscape. In this situation the density
of inactive particles is expected to decay exponentially, $n_l \sim
\exp(-l/l_w)$, where $l$ measures the distance from the condensate and
$l_w$ is the typical width of the cloud. We estimate $l_w$ by the
following consideration.  The correlated cluster at the condensate has
the largest size among the clusters and its value, $\xi_L$, follows
from extreme value statistics. Using the distribution function of
cluster sizes below Eq.~(\ref{xi}) we obtain for its typical value:
$\xi_L \sim \xi \ln L $.  At this distance, $l=\xi_L$, the typical
value of the weight is $g_l=O(1)$, and with Eqs.~(\ref{nav}),
(\ref{geom}), and (\ref{g_L}) we obtain $n_l\sim L^{-z} \sim
\exp(-\xi_L/l_w)$. Consequently
\be
l_w \sim \frac{\xi}{z}=\xi_{typ} \sim \delta^{-1}\;,
\label{l_w}
\ee
where we have made use of the scaling relation\cite{fisher99,im}:
$\xi/\xi_{typ}=z$, where $\xi_{typ}$ is the typical
correlation length. The small $\delta$ behavior in the last equation
follows directly from Eqs.(\ref{xi}) and (\ref{z_small}). We have
obtained thus the result that the typical width of the cloud of
inactive particles is measured by the typical correlation length of
the biased Sinai walker.

\begin{figure}
\includegraphics[width=0.8\linewidth]{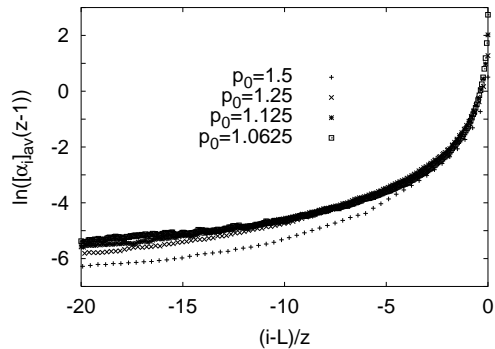}
\caption{\label{pfg} Average of the occupation probability, $\alpha_i$, which corresponds to
the typical density of the cloud. The data were obtained by solving
Eq.(\ref{finite}) numerically for $L=N=2048$ and the disorder average was
performed over 500000 samples.
We used the uniform distribution in Eq.(\ref{uniform}) for different values of the
asymmetry parameter, $\delta$, and thus for the dynamical exponent, $z>1$.
The data are rescaled according to (\ref{alpha1}) and using the length scale
in (\ref{l_w}) with (\ref{z}).
  }  
\end{figure}

The typical density profile of inactive particles has the same scaling
behavior as the typical and the average value of the occupation
number, $\alpha_i$, as defined in Eq.~(\ref{geom}). For this latter
quantity we have checked numerically the scaling form in
Eq.~(\ref{l_w}) for different values of $\delta$ and satisfactory
agreement is found, see Fig.~\ref{pfg}.

%%%%%%%%%%%%%%%%%%%%%%%%%%%%%%%%%%%%%%%%%%%%%%%%%%%%%%%%%%%%%%%%%%%%%%%%%%%
%%%%%%%%%%%%%%%%%%%%%%%%%%%%%%%%%%%%%%%%%%%%%%%%%%%%%%%%%%%%%%%%%%%%%%%%%%%

\subsubsection{Average number of inactive particles}
\label{sec_inactive}

The typical value of inactive particles is finite, however in rare
realizations it is possible to find even a macroscopic number of inactive
particles. This is due to the fact that the leading value of the
weights in Eq.~(\ref{g}), $g_L$, and the subleading value,
$g_{\tilde{l}} \equiv g$
can be arbitrarily close to each other.
The average value of $[\langle n_{\tilde{l}} \rangle ]_{\rm av}\approx
N_{ia}$ is dominated by such rare realizations, in which both $\langle n_L \rangle$ and
$\langle n_{\tilde{l}} \rangle$ are macroscopic, i.e. they are of $\mathcal{O}(N)$.
However, as we checked numerically, the contribution to the average value of such samples in which more than
two sites have macroscopic occupation is negligible. Therefore in the subset of rare
realizations we should consider only two active sites, $L$ and $\tilde{l}$,
and the unnormalized weight that the subleading site contains $n_{\tilde{l}}=n$
particles is in leading order 
proportional to $g^{n}g_L^{N-n}= \alpha^{n}g_L^N$.
The distribution of $n$ is thus approximately 
$P_{\alpha}(n)=(1-\alpha)/(1-\alpha^{N+1})\alpha^{n}$ for $\alpha<1$ and $P_{\alpha}(n)=1/(N+1)$,
for $\alpha=1$\cite{current}. Thus we obtain for the expectation value:
\be
\langle n \rangle (\alpha)={\alpha\over 
1-\alpha}-(N+1){ \alpha^{N+1}\over 1-\alpha^{N+1}},\quad \alpha<1\;,
\label{corr}
\ee
and $\langle n \rangle=N/2$ for $\alpha=1$. Using the distribution function, $\rho(\alpha)$ we
can average over the rare realizations:  
$[\langle n_{\tilde{l}} \rangle ]_{\rm av}=
\int_0^1 \langle n \rangle (\alpha)\rho(\alpha) {\rm d}\alpha$, which is dominated by the
contribution as $\alpha \to 1$. Keeping in mind that the maximal value of $\langle n \rangle$ is
$N/2$, we can write
\be
 N_{ia}\approx [\langle n_{\tilde{l}} \rangle ]_{\rm av} \approx
\int_0^{1-2/N} \frac{\alpha}{1-\alpha}\rho(\alpha) {\rm d}\alpha \sim \rho(1)
 \ln N.
\ee
provided that the distribution $\rho(\alpha)$ has a finite limiting 
value at $\alpha \to 1$. As we have
checked numerically this is indeed the case, both for
the unbiased and  the biased (asymmetric) models. Consequently
the average number of inactive particles is logarithmically divergent, although its typical value
is finite. Furthermore the ratio between the average numbers of the active and inactive particles,
$N_a/N_{ia}$, tends to zero for $z>1$ and tends to infinity, for $z<1$.

\subsubsection{Average density profile of inactive particles}
\label{sec_density}

\begin{figure}
\includegraphics[width=0.45\linewidth]{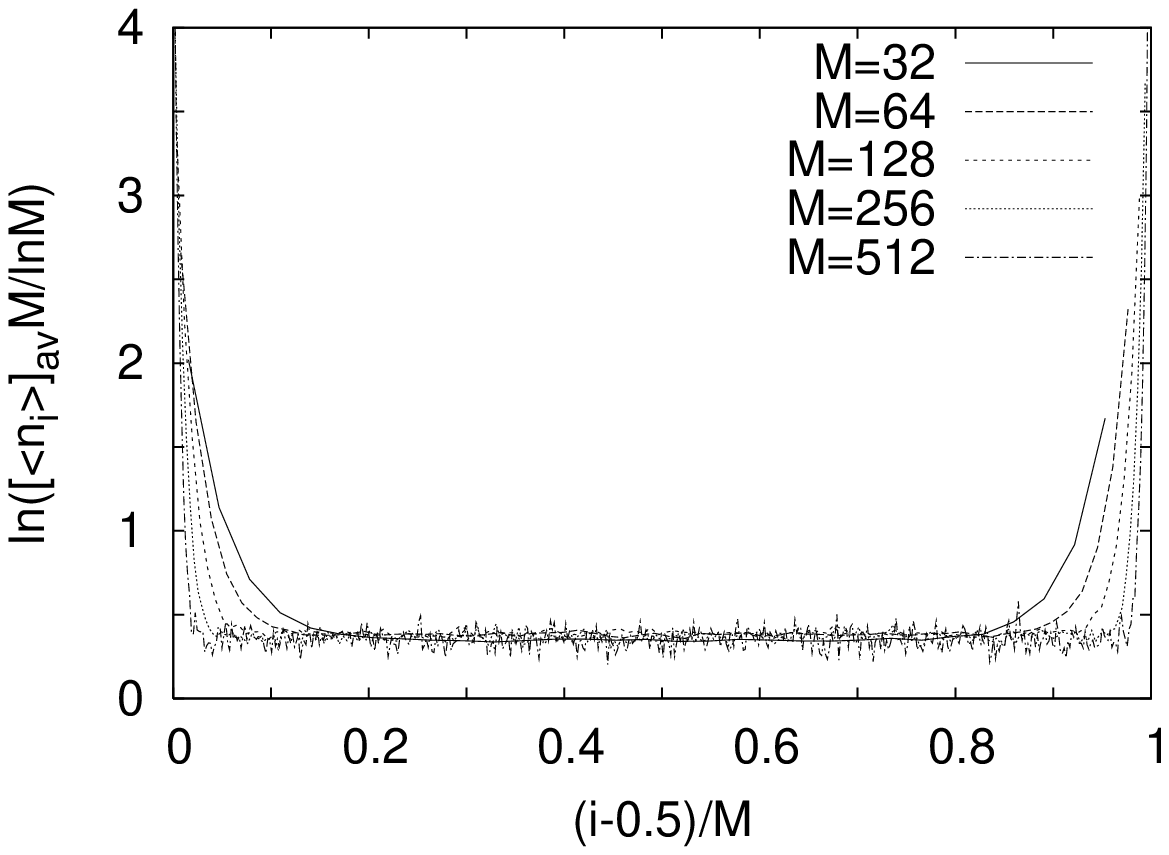}
\includegraphics[width=0.45\linewidth]{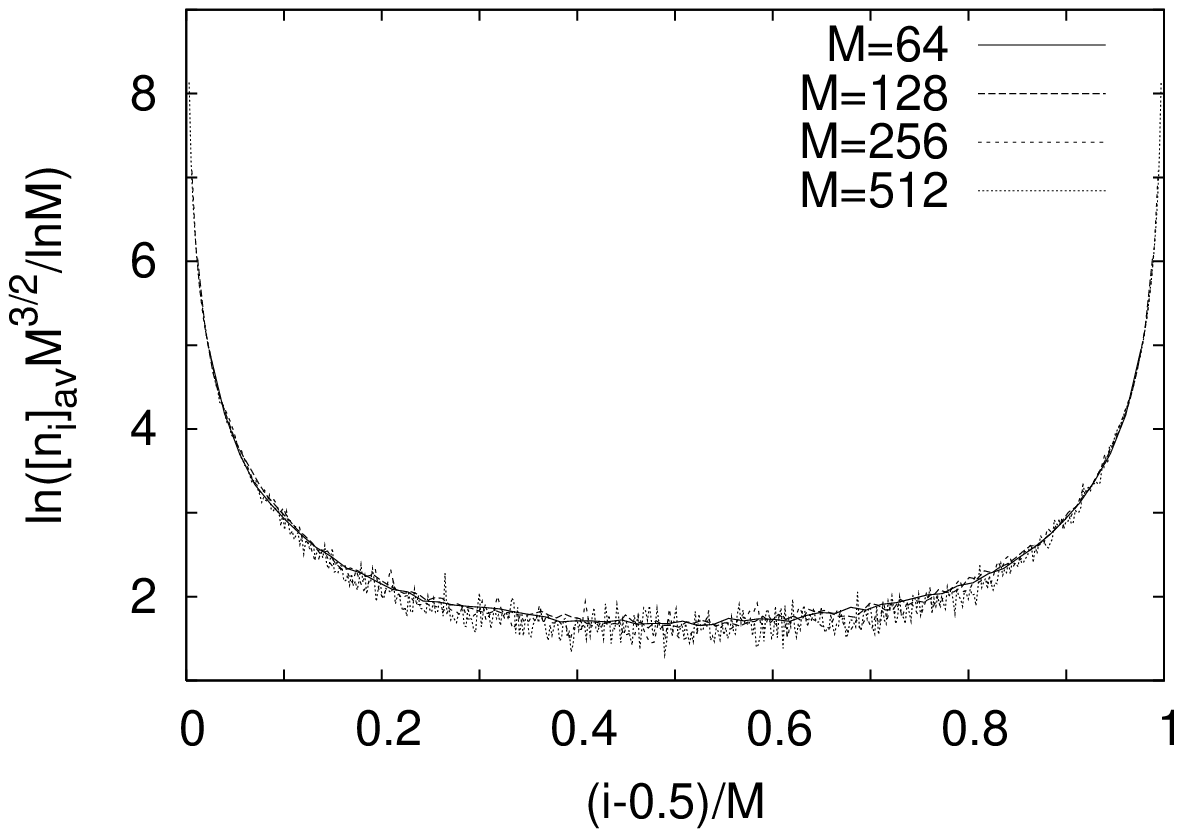}
\caption{\label{pf} Average density profiles of inactive particles
 for different system sizes, obtained by solving
Eq.(\ref{finite}) numerically. The density was $\rho =1$ and average
 was performed 
 over 500000 samples. 
Left: asymmetric model with the uniform distribution, $p_0=3$. The profile is constant outside
the renormalized cluster of the condensate, the size of which is finite.
Right: unbiased model, $p_0=1$, rescaled according to (\ref{scaling}).}  
\end{figure}

The average density profile of inactive particles, $[\langle n_l
\rangle]_{\rm av}$, is proportional to $P(l)$, which is the probability
density, that the subleading, almost degenerate $g \approx g_L$ is located at 
$l=\tilde{l}$. In the asymmetric model $\tilde{l}$ can be either at the
renormalized cluster of the condensate or at any other cluster. 
In the second case $P(l)$ is a constant, which happens for $l > \xi$. 
This behavior is illustrated in the left part of Fig.~\ref{pf}.

In the unbiased model, as can be seen from Eq.~(\ref{xi}) the width
of the cluster diverges. Therefore $\tilde{l}$ and $L$ are always
in the same cluster, and as a consequence the probability distribution is
scale-free and a function of $l/L$. 
The form of the probability distribution for $l \ll L$ 
can be obtained from the random walk picture in Fig.~\ref{rw1}.
The rare event for this process is represented by a landscape which is
drawn by an unbiased ($\delta=0$) random walker, which starts at one
minimum and after $l$ steps arrives to a degenerate second minimum. This
means that the random walker is surviving (since it does not cross its
starting position) and returns to its origin. The fraction of such
random walks is given by\cite{ir}: $P(l) \sim l^{-3/2}$, $l \ll L$,
consequently $P(l)=L^{-3/2} \tilde{P}(l/L)$, where $\tilde{P}(x)$ is a
smooth scaling function, which behaves for small $x$ as $\tilde{P}(x)
\sim x^{-3/2}$. With this prerequisite we obtain for the average density
profile of inactive particles for a given density $\rho=N/L>0$:
\be 
[\langle n_l\rangle ]_{av}(L)=\ln (L)L^{-3/2}\tilde{P}(l/L).
\label{scaling}
\ee
Results of numerical calculations are presented in the right part of
Fig.~\ref{pf}, which are in excellent agreement with the theoretical 
predictions.

 %%%%%%%%%%%%%%%%%%%%%%%%%%%%%%%%%%%%%%%%%%%%%%%%%%%%%%%%%%%%%%%%%%%%%%%%%%%
%%%%%%%%%%%%%%%%%%%%%%%%%%%%%%%%%%%%%%%%%%%%%%%%%%%%%%%%%%%%%%%%%%%%%%%%%%%

%%%%%%%%%%%%%%%%%%%%%%%%%%%%%%%%%%%%%%%%%%%%%%%%%%%%%%%%%%%%%%%%%%%%%%%%%%%
%%%%%%%%%%%%%%%%%%%%%%%%%%%%%%%%%%%%%%%%%%%%%%%%%%%%%%%%%%%%%%%%%%%%%%%%%%%

\section{Coarsening}
\label{sec_coarsening}
In the stationary state of the process typically almost all particles
occupy single site. If initially the particles are uniformly 
distributed on the lattice, the system undergoes a coarsening process,
meaning that the number of particles in the condensate and the typical
size of empty regions is growing as time elapses. To be specific we
consider finite $L$ and $N$ and define the length scale as\cite{kf} $l(t)\equiv
\sigma^2(t)=[{1\over L}\sum_{i=1}^L\langle n_i^2\rangle (t)]_{av}$
, which characterizes the typical number of particles at 
sites with a non-microscopic $n_i$, or equivalently the typical 
distance between
these sites-  In the RG procedure
the latter corresponds to the typical cluster sizes (defined
as the sum of the original links) at the energy scale $\Omega$.
The growth rate of the typical distance, ${\rm d} l /{\rm d} t$,
is proportional to the typical current at that energy scale.
This  leads to the differential equation:
\be
\frac{{\rm d} l}{{\rm d} t} \sim J_l\;.
\label{l_t}
\ee
At the Griffiths phase we have in (\ref{jgriffiths}), $J_l\sim l^{-z}$,
thus the solution of Eq.~(\ref{l_t}) is given by:
\be 
l \sim t^{1/\zeta},\quad \zeta=z+1\;.
\label{dyn}
\ee

Here $\zeta$ is the dynamical coarsening exponent. Note that $\zeta$ is exactly
known through Eq.(\ref{z}) and it is a continuous function of
the parameters appearing in the distributions of hop rates.
Close to the critical point $\zeta$ becomes universal, 
depending only on the control parameter: 
$\zeta\approx 1/(2\delta )$, see Eq.(\ref{z_small}). 
We note that the same result can be obtained by referring to the result of the
RG procedure.  As shown in subsection \ref{sec_asym} for $\Omega <
\Omega_{\xi}$ the ZRP is transformed to a totally asymmetric ASEP with a rate
distribution in Eq.~(\ref{Pp}). The coarsening of the 
totally asymmetric ASEP
has been analyzed in Refs.~[\onlinecite{kf,jainbarma,bennaim}] 
with the result given in Eq.~(\ref{dyn}).
 
At the critical point  $\zeta$ diverges and the coarsening is ultra-slow
and strictly at $\delta = 0$ the length scale grows
anomalously (logarithmically) with $t$. 
This type of growth is the so called  {\it anomalous coarsening} 
\cite{evanscoars},  
which can be observed e.g. in spin glasses.  

\begin{figure}
\includegraphics[width=0.45\linewidth]{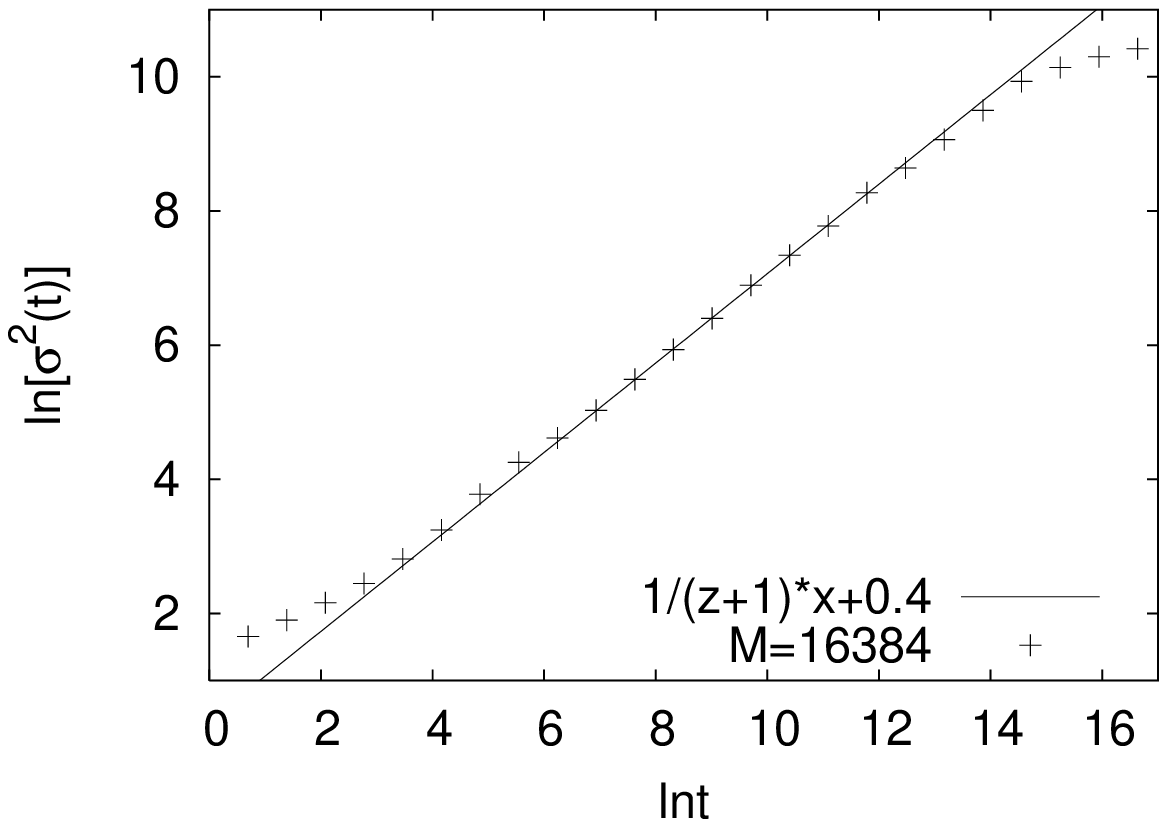}
\includegraphics[width=0.45\linewidth]{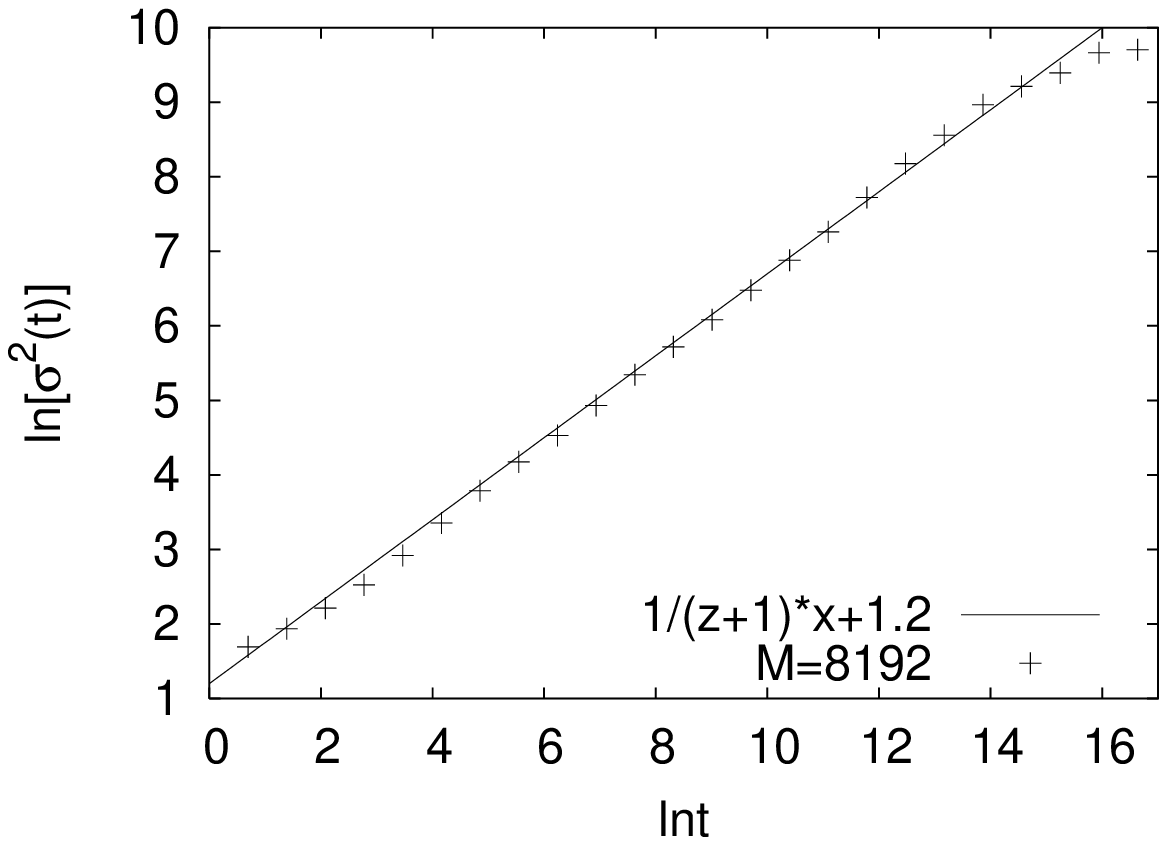}
\caption{\label{fig20} 
 Time dependence of the coarsening length scale, $l$, in the
 Griffiths phase in a log-log plot. We used the bimodal
 distribution (\ref{bimodal}) with c=0.2, r=0.5 where $z=0.5$ (left)
 and  c=0.3, r=0.5 where $z\approx 0.818$ (right), having a density
 $\rho =3$, and disorder average was
 performed over a few hundred samples. The slope of the straight line
 is the theoretical result in Eq.(\ref{dyn}): $1/(1+z)$.}  
\end{figure}

\begin{figure}
\includegraphics[width=0.45\linewidth]{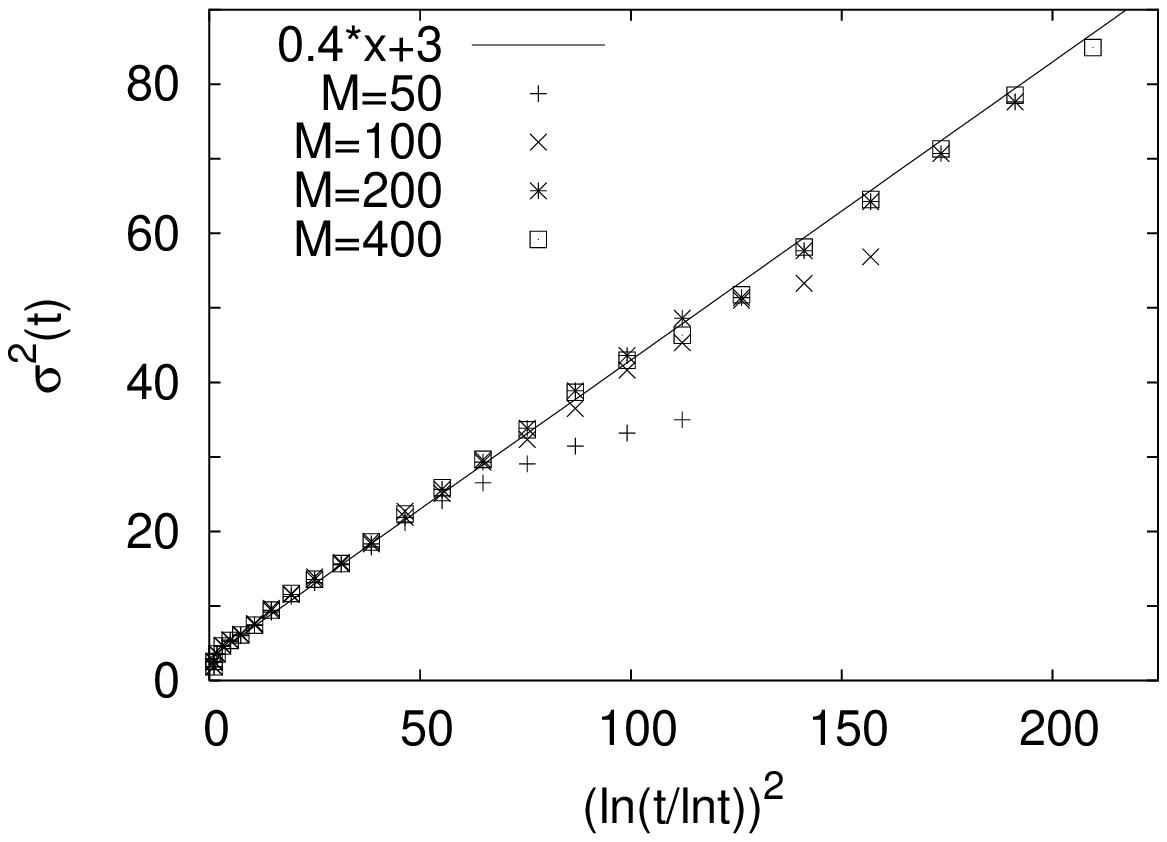}
\includegraphics[width=0.45\linewidth]{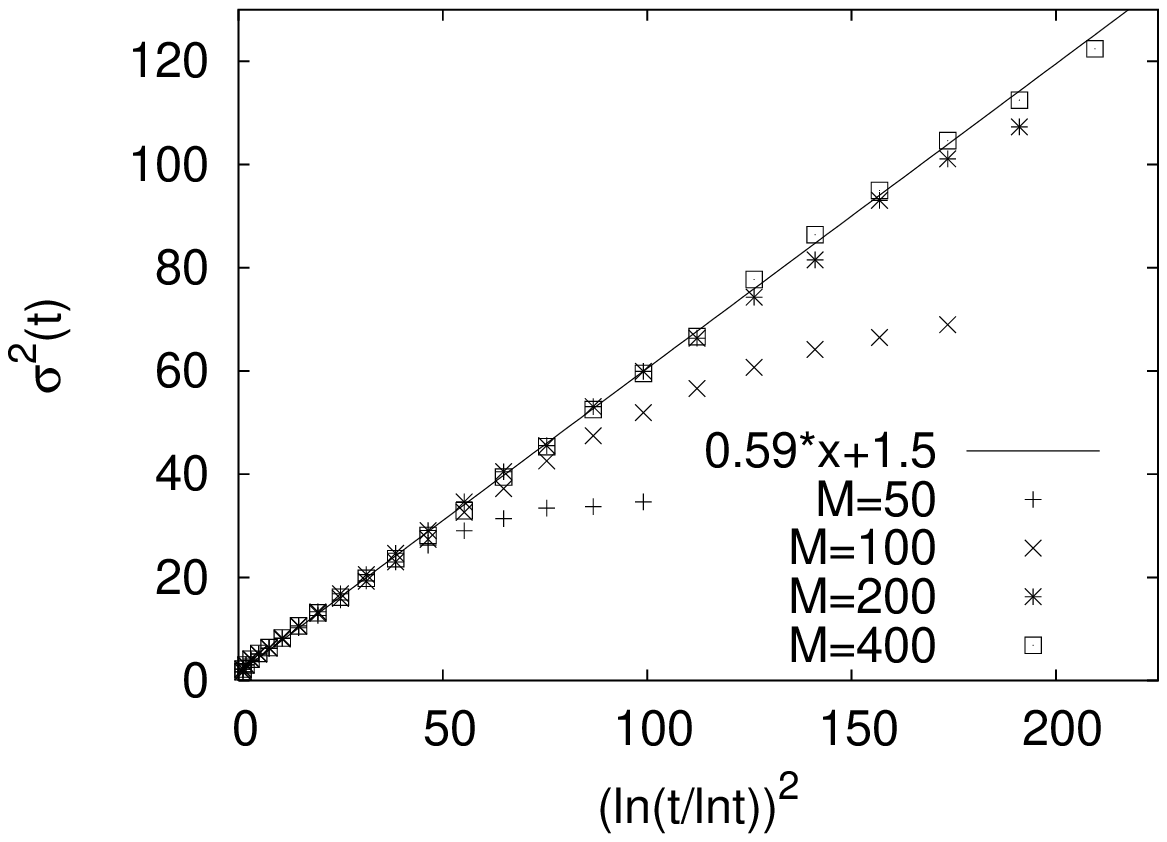}
\caption{\label{fig21} 
Time dependence of the coarsening length scale, $l$, for the unbiased
model using the theoretical combination in Eq.~(\ref{log}). We used the bimodal
 distribution (\ref{bimodal})) with
c=0.5, r=0.1 (left) and  c=0.5, r=0.2 (right), having a density
 $\rho =3$, and disorder average was
 performed over a few hundred samples.}  
\end{figure}

The asymptotic time dependence of $l$ can be obtained by inserting
the scaling form of the typical current through an empty region 
of size $l$ into the differential equation in Eq.~(\ref{l_t}). The 
scaling form is according (\ref{jcrit}) given by $J_l\sim e^{-c l^{1/2}}$.
Eq.~(\ref{jcrit}) has the asymptotic (large $t$) solution $l^{1/2}e^{{\rm c}
l^{1/2}}\sim t$, 
thus the growth of the length scale is logarithmically slow
\be 
l \sim \left[\ln\left({t\over \ln t}\right) \right]^2.
\label{log}
\ee
Results of numerical simulations are presented in Fig. \ref{fig20} for
the Griffiths phase (biased model) and in Fig. \ref{fig21} for the unbiased model. The
agreement with the theoretical results is satisfactory.

 %%%%%%%%%%%%%%%%%%%%%%%%%%%%%%%%%%%%%%%%%%%%%%%%%%%%%%%%%%%%%%%%%%%%%%%%%%%
%%%%%%%%%%%%%%%%%%%%%%%%%%%%%%%%%%%%%%%%%%%%%%%%%%%%%%%%%%%%%%%%%%%%%%%%%%%

%%%%%%%%%%%%%%%%%%%%%%%%%%%%%%%%%%%%%%%%%%%%%%%%%%%%%%%%%%%%%%%%%%%%%%%%%%%
%%%%%%%%%%%%%%%%%%%%%%%%%%%%%%%%%%%%%%%%%%%%%%%%%%%%%%%%%%%%%%%%%%%%%%%%%%%

At the boundary of the Griffiths phase, where $z=0$, 
the coarsening is controlled 
by the behavior of the forward rate distribution at $p\to 0$. 
To see this, let us consider an arbitrary forward rate distribution for which
$\rho (p)\sim p^{-1+1/z_0}$~~($z_0>0$) as $p\to 0$, and an arbitrary backward rate distribution, 
%e.g. $q_i=q_0$ for all $i$. 
such that $\delta >0$.
Analyzing (\ref{z}) it is easy to show that $\zeta > z_0+1$ in the
whole Griffiths phase $\delta >0$, and at the boundary 
\be 
\lim_{\delta\to\infty}\zeta (\delta )= z_0+1,
\ee
which is the coarsening exponent of a totally
asymmetric process obtained from the original one by setting 
the backward rates to zero. 
Thus, as expected, the presence of nonzero backward rates slows down 
the coarsening in so far as the dynamical exponent is larger in the
whole Griffiths phase than that of 
the totally asymmetric process with zero backward rates, and the latter 
value is recovered at the boundary of the Griffiths phase.

%%%%%%%%%%%%%%%%%%%%%%%%%%%%%%%%%%%%%%%%%%%%%%%%%%%%%%%%%%%%%%%%%%%%%%%
%%%%%%%%%%%%%%%%%%%%%%%%%%%%%%%%%%%%%%%%%%%%%%%%%%%%%%%%%%%%%%%%%%%%%%%
\section{Conclusion}

In this paper the prototype of a diffusive many particle system, the
zero range process, is studied in one dimension, in the presence of
quenched disorder. This model system is of large importance since its
stationary state is in a product form therefore can be constructed
analytically. At the same time there exists a wide range of 
different realisations of the ZRP, and therefore it can be applied 
to various problems of stochastic transport.
We have analyzed the properties of the steady state of
the disordered system and obtained many exact results. These results
are expected to be generic for random, driven non-equilibrium systems
and therefore can be useful to analyze more complicated,
non-integrable processes, too. It also important to notice 
that our results are directly related to another archetypical model
of non-equilibrium transport, i.e. the
ASEP with particle-wise disorder.

\subsection{Types of the transport}

The transport in the system is related to the form of the 
hop rate distribution and characterized by the value of the dynamical
exponent $z$ defined in Eq.~(\ref{z}).  We can distinguish four main 
types of transport in the random ZRP.

\subsubsection{$z=0$}

In this case the maximal value of the backward hop rates, $q_{max}$,
is smaller than the minimal value of the forward hop rates, $p_{min} >
q_{max}$. This situation is equivalent to the totally asymmetric ZRP,
which has been previously analyzed in the literature. In this case
there is a finite stationary current in the system, such that the number
of active particles (which carry the current) is $N_a \sim L$ and the
particles have a finite velocity, $v=O(1)$. Condensation of 
particles at a particular site is not generic, but appears only
for hop rate distributions that are vanishing fast enough at the 
lower cut-off $c>0$ (see \cite{krug} for details).  

\subsubsection{$0<z<1$}

In this case $q_{max}>q_{min}$ and the current is vanishing in the
thermodynamic limit as $J \sim L^{-z}$. The transport in the system is
effected by $N_a \sim L^{1-z}$ active particles, which move with a finite
velocity, $v=O(1)$. Note that in the system there are infinitely many
active particles, but their density is zero.

\subsubsection{$1<z<\infty$}

The current in the system  is vanishing according to $J \sim L^{-z}$, 
and the
transport is effected by $N_a=O(1)$ active particles, which have a
vanishing velocity. Thus the transport is realized by the anomalous
diffusion of a few biased random walker.

\subsubsection{Unbiased ZRP: $z \to \infty$}

The average current is zero and the fluctuations of the accumulated
displacement are due to a few Sinai walkers and given in
Eq.~(\ref{sinai}).

\subsection{Problem with self averaging}

In the partially asymmetric random ZRP with $z>0$ the number of active
particles, $N_a$, and thus the transport is self averaging, i.e. the
typical and the average values have the same scaling behavior. The
inactive particles, however, which are outside the condensate but do
not contribute to the transport, have different properties. Their
typical number is $O(1)$ whereas their average diverges as $\ln
N$. This latter value is dominated by rare realizations in which there
is a second site with macroscopic number of particles.

\subsection{Relation with the totally asymmetric ZRP}

The RG framework used in this paper has revealed a
relation between the partially asymmetric ZRP with arbitrary type of
initial disorder and the totally asymmetric ZRP with a hop 
rate distribution,
which has a vanishing power-law tail. Indeed, during renormalization
if the typical size of the renormalized new clusters becomes larger
than the correlation length the new effective particles perform a totally
asymmetric motion with a power-law distribution of the hop rates given
in Eq.~(\ref{Pp}). The exponent of this distribution is related to the
form of the distribution in the original (i.e. partially asymmetric)
model, see in Eq.~(\ref{z}). In this way the problem which has been
throughly studied for the random totally asymmetric ZRP appears
naturally as the fixed point problem of the partially asymmetric ZRP.
The power-law exponent then appears in a self-organized fashion. In
the terminology of random systems both problems have the same type of
{\it strong disorder fixed point}. We note that a similar relation is
encountered between the biased Sinai walk and the directed trap model\cite{im}.
On the other hand for the unbiased ZRP the
singular behavior is governed by a so called {\it infinite disorder fixed point}.

\subsection{Condensation transition for limited number of particles}

Throughout the analysis of
the partially asymmetric model we considered exclusively the 
case of a finite density, $\rho=N/L>0$. In this case  there is 
always a condensate present, where one typically finds 
a finite fraction of the particles. However, if the number 
of particles scales as $N \sim
L^{\omega}$, $0<\omega<1$, one expects that the condensate will
disappear for sufficiently small values of $\omega$. Indeed, the 
transport in the system can involve $N_a \sim L^{1-z}$ particles, for $0<z<1$,
consequently for $\omega<1-z$ all the particles carry current and the
condensate is absent in the system.
At the borderline case, with $N=A L^{1-z}$ one expects
that the fraction of particles in the condensate varies with $A$ and
that there is a condensation transition possibly at a finite value of
$A=A_c=O(1)$. First numerical results support this scenario of the condensation 
transition, but a more detailed analysis of the phase transition,
which is due to the slow dynamics of the process quite involved,
is not yet completed. We are going to clarify this phenomena in 
a separate work.

\subsection{Higher dimensions}

The ZRP process has the same type of factorized steady state for any
type of lattices, thus also in higher dimensions. Also the stationary
weights involve site dependent quantities, $g_i$, which are the
stationary weights of the random walk on the given lattice. The
renormalization described in Sec.\ref{sec_RG} can be implemented
numerically in this case. Similar calculations have already been
performed for random quantum systems\cite{im}. During the RG procedure one
obtains large effective clusters having a complicated topology. One
might ask the question whether the strong and infinite disorder nature of
the fixed points as obtained in 1d remains also in higher dimensions.
The answer to this question is negative. We know that the underlying
random walk process has an upper critical dimension, $d_u=2$, so that
the Gaussian nature of the random walk remains\cite{luck} (in $d=2$ with
logarithmic corrections) even in the presence of quenched disorder.
The same type of irrelevance of disorder is expected to hold for the
ZRP, too.

%%%%%%%%%%%%%%%%%%%%%%%%%%%%%%%%%%%%%%%%%%%%%%%%%%%%%%%%%%%%%%%%%%%%%%%%%%%
\section{Acknowledgments}

R.J. is indebted to H. Rieger for stimulating discussions. 
R.J. and L.S. acknowledge support by the Deutsche
Forschungsgemeinschaft under grant No. SA864/2-1. 
This work has been
supported by a German-Hungarian exchange program (DAAD-M\"OB), by the
Hungarian National Research Fund under grant No OTKA TO34138, TO37323,
TO48721, MO45596 and M36803.

%%%%%%%%%%%%%%%%%%%%%%%%%%%%%%%%%%%%%%%%%%%%%%%%%%%%%%%%%%%%%%%%%%%%%%%
%%%%%%%%%%%%%%%%%%%%%%%%%%%%%%%%%%%%%%%%%%%%%%%%%%%%%%%%%%%%%%%%%%%%%%%

\appendix

\section{Mapping to the RTFIC, the RG equations and their fixed point solution}

Renormalization of the one-dimensional ZRP with quenched disorder
as described in Sec.~\ref{sec_RG} is equivalent to the same procedure of a
random transverse-field Ising chain (RTFIC) described by the Hamiltonian:
\be
H=-\sum_{l=1}^{L} J_i
\sigma^x_{l}\sigma^x_{l+1}-\sum_{l=1}^{L}h_l\sigma^z_{l},
\label{Ising}
\ee
where $\sigma^{x,z}_l$ are Pauli spin operators at site $l$, and the 
couplings, $J_l$, and the transverse fields, $h_l$, are  
quenched random variables. During renormalization the largest term in
the Hamiltonian, $\Omega=\max(\{J_i\},\{h_i\})$ is successively eliminated.
For example, if the largest term is a coupling, $\Omega=J_i$, the connected two sites, $(i,i+1)$,
flip coherently and form an effective two-site cluster. The renormalized value
of the transverse field acting on the spin cluster is given by a second-order perturbation
calculation as:
\be
\tilde{h} \approx \frac{h_i h_{i+1}}{\Omega}\;.
\label{htilde}
\ee
Decimating a strong transverse field, $\Omega=h_i$, will result in decimating out site, $i$,
and generating a new coupling, $\tilde{J}$, between the remaining sites, $i-1$ and $i+1$. The
value of $\tilde{J}$ follows from self-duality, in Eq.~(\ref{htilde}) one should replace $h_i$
by $J_i$. Comparing the decimation rules for the ZRP in Eq.~(\ref{back}) to that for
the RTFIC in Eq.~(\ref{htilde}) one observes a complete agreement with the correspondences:
$p_i \leftrightarrow J_i$ and $q_i \leftrightarrow h_i$. Now properties of the RG flow
for the ZRP can be obtained from the equivalent expressions for the RTFIC.

For example the probability distribution of the backward, $R(q,\Omega)$, and forward,
$P(p,\Omega)$, hop rates satisfy the set of integral-differential equations,
\beqn 
{{\rm d} R(q,\Omega) \over {\rm d}
\Omega} &=& R(q,\Omega)[P(\Omega,\Omega)-R(\Omega,\Omega)]\\ 
& - & P(\Omega,\Omega) \int_{q}^{\Omega} {\rm d} q' R(q',\Omega) R({q \Omega \over q'},\Omega)
{\Omega \over q'} \nonumber \\
{{\rm d} P(p,\Omega) \over {\rm d}
\Omega}&=&P(p,\Omega)[R(\Omega,\Omega)-P(\Omega,\Omega)]\\
 & - &R(\Omega,\Omega) \int_{p}^{\Omega} {\rm d} p' P(p',\Omega) P({p \Omega \over p'},\Omega)
{\Omega \over q'}, \nonumber
\eeqn
having a solution at the fixed point:
\beqn 
P_0(p,\Omega)=\frac{p_0(\Omega)}{\Omega}\left(\frac{\Omega}{p}\right)^{1-p_0(\Omega )},  \nonumber \\
R_0(q,\Omega)=\frac{r_0(\Omega)}{\Omega}\left(\frac{\Omega}{q}\right)^{1-r_0(\Omega )},
\label{sol}
\eeqn
with $0<p,q\le \Omega$.
The value of the exponents, $p_0(\Omega)$ and $r_0(\Omega)$, depend on original distributions
and therefore on the value of the control parameter, $\delta$. The specific values are given in
Sec.\ref{sec_RG}.

\section{Average of the occupation probability}

Here we start with the asymptotic distribution of the (uncorrelated) $g_i$ weights in
Eq.~(\ref{kesten}) with the condition that the largest value is fixed, $g_L=G$,
so that
\be
\rho(g|G)dg={1/zg^{-1/z-1}\over 1-G^{-1/z}}\Theta(G-g).
\ee
From this we obtain for the distribution of the occupation probabilities, $\alpha_i=g_i/g_L$, for large $L$ as  
\be
\rho(\alpha)d\alpha\approx {1\over z}{1\over \alpha}\left[{1\over L\alpha^{1/z}}-
e^{- L\alpha^{1/z}}\left(1+{1\over L\alpha^{1/z}} \right) \right]d\alpha.
\ee
The average of $\alpha$ is therefore given by:
\beqn 
[\alpha]_{av}&=&\int_0^1\alpha\rho(\alpha)d\alpha \nonumber \\
&\approx &
L^{-z}\int_0^L\left[{1\over y}-e^{-y}\left(1+{1\over y}\right) \right]
y^{-1+z}dy \ .
\eeqn
In leading order of $L$ the second term in the bracket can be neglected and we arrive at:
\beqn
[\alpha]_{\rm av} &\approx& L^{-z}\left({\rm const}+{L^{z-1}\over z-1}\right)\sim
L^{-z},\quad z<1, \nonumber \\
\left[\alpha\right]_{\rm av} &\approx& L^{-1}\left({\rm const}+\ln L\right)\sim
L^{-1}\ln L \quad z=1,\nonumber \\
\left[\alpha\right]_{\rm av}&\approx& L^{-z}\left({\rm const}+{L^{z-1}\over z-1}\right)\sim
\frac{L^{-1}}{z-1} \quad z>1. \label{alpha1}
\eeqn
Thus, as far as $z>1$ the occupation probability is
non-self-averaging, and the decay exponent is becoming
one independent of $\delta$.


\begin{thebibliography}{}

\bibitem{krug} J. Krug, Braz. J. Phys. {\bf 30}, 97 (2000).

\bibitem{im} F. Igl\'oi and C. Monthus, Physics Reports {\bf 412}, 277, (2005),
preprint cond-mat/0502448.

\bibitem{barma} G. Tripathy and M. Barma, Phys. Rev. Lett. {\bf 78}, 3039 (1997); Phys. Rev. E {\bf 58}, 1911 (1998.)

\bibitem{janowsky} S.\ A.\ Janowsky and J.L. Lebowitz, Phys. Rev A {\bf 45}, 618 (1992); J. Stat. Phys. {\bf 77}, 35 (1994).

\bibitem{sinai} For a review see: J.P. Bouchaud and A. Georges, Phys. Rep. {\bf 195}, 127 (1990).

\bibitem{kln04} Y.~Kafri, D.K.~Lubensky, and D.R. Nelson, 
Biophys. J. {\bf 86}, 3373 (2004).

\bibitem{kln05} Y.~Kafri, D.K.~Lubensky, and D.R. Nelson, 
Phys.~Rev. E {\bf 71}, 041906 (2005).

\bibitem{evansreview} For a review see: M.R. Evans and T. Hanney,
  preprint cond-mat/0501338.

\bibitem{spitzer} F. Spitzer (1970) Advances in Math. {\bf 5} 246.

\bibitem{chowd} D. Chowdury, L.~Santen, A.~Schadschneider, Phys. Rep. {\bf 329}, 199 (2000).

\bibitem{jainbarma} K. Jain and M. Barma, Phys. Rev. Lett. {\bf 91}, 135701 (2003).

\bibitem{liggett} T.M. Liggett {\it Stochastic interacting systems:
  contact, voter, and exclusion processes}, (Berlin, Springer, 1999).

\bibitem{hinrichsen} H. Hinrichsen, Adv.~Phys.~{\bf 49}, 815 (2000).

\bibitem{schutzreview}  G.M. Sch\"utz, in {\it Phase Transitions and
Critical Phenomena}, vol. 19, Eds. C. Domb and J.L.  Lebowitz (Academic Press, San Diego, 2001).

\bibitem{kf} J. Krug and P. A. Ferrari, J. Phys. A {\bf 29}, L465 (1996).

\bibitem{evans} M. R. Evans, Europhys. Lett. {\bf 36}, 13 (1996).

\bibitem{jsi} R. Juh\'asz, L. Santen and F. Igl\'oi,
  Phys. Rev. Lett. {\bf 94}, 010601 (2005).

\bibitem{derrida} C. Enaud and B. Derrida, Europhys. Lett. {\bf 66}, 83 (2004).

\bibitem{stinchcombe} R.J. Harris and R.B. Stinchcombe, Phys. Rev. E {\bf 70} 016108(E) (2004).

\bibitem{MDH} S.K. Ma, C. Dasgupta, and C.-K. Hu, \prl {\bf 43}, 1434 (1979);
C. Dasgupta and S.K. Ma, Phys. Rev. B {\bf 22}, 1305 (1980).

\bibitem{DF} D.S. Fisher, Phys. Rev. Lett. {\bf 69}, 534 (1992); Phys. Rev. B {\bf 51}, 6411 (1995).

\bibitem{fisherxx} D.S. Fisher, Phys. Rev. B {\bf 50}, 3799 (1995).

\bibitem{RGsinai} D.S. Fisher, P. Le Doussal, and C. Monthus, Phys. Rev. E{\bf 59} 4795 (1999).

\bibitem{hiv} J. Hooyberghs, F. Igl\'{o}i, and C. Vanderzande, Phys. Rev. Lett.  {\bf 90}, 100601 (2003);
Phys. Rev. E {\bf 69}, 066140 (2004).

\bibitem{kafri} Y. Kafri, E. Levine, D. Mukamel, G.M. Sch\"utz and
  J. T\"or\"ok, Phys. Rev. Lett. {\bf 89}, 035702 (2002).

\bibitem{griffiths} R.B. Griffiths, Phys. Rev. Lett. {\bf 23}, 17 (1969).

\bibitem{kesten} H. Kesten, Acta Math. {\bf 131}, 298 (1973);
B. Derrida and H. Hilhorst, J. Phys. A{\bf 16}, 2641 (1983); C. de
Calan, J. M. Luck, T. M. Nieuwenhuizen and D. Petritis, J. Phys. A {\bf 18}, 501 (1985).

\bibitem{galambos} J. Galambos, {\it The Asymptotic Theory of Extreme
Order Statistics} (John Wiley and Sons, New York, 1978).

\bibitem{ir} F. Igl\'oi and H. Rieger,  Phys. Rev. B {\bf 57}, 11404 (1998).

\bibitem{i02} F. Igl\'oi, Phys. Rev. B{\bf 65}, 064416 (2002).

\bibitem{sinai1} Y.A.G. Sinai, Theor. Prob. Appl. {\bf 27}, 256 (1982).

\bibitem{fisher99}
D.S. Fisher, Physica A {\bf 263}, 222 (1999).

\bibitem{current} In the framework of this simplified picture we
  obtain for the finite $N$ correction to the current:
  $J_L/J_{L,N}\approx 1+P_{\alpha}(N)$.

\bibitem{bennaim} E. Ben-Naim, P.L. Krapivsky, and S. Redner,
  Phys. Rev. E {\bf 50}, 822 (1994).
 
\bibitem{evanscoars} M.R. Evans, J. Phys: Condens. Matter {\bf 14},
  1397 (2002).

\bibitem{luck}
J.M. Luck, Nucl. Phys. B {\bf 225}, 169 (1983); D.S. Fisher, Phys. Rev. A {\bf 30}, 960 (1984).

\end{thebibliography}
\end{document}